# Angle Dependent Van Hove Singularities in Slightly Twisted Graphene Bilayer

Wei Yan[1,§], Mengxi Liu[2,§], Rui-Fen Dou[1], Lan Meng[1], Lei Feng[1], Zhao-Dong Chu[1], Yanfeng Zhang[2,3,*], Zhongfan Liu[2], Jia-Cai Nie[1], and Lin He[1,*]

[1] Department of Physics, Beijing Normal University, Beijing, 100875, People's Republic of China
[2] Center for Nanochemistry (CNC), College of Chemistry and Molecular Engineering, Peking University, Beijing 100871, People's Republic of China
[3] Department of Materials Science and Engineering, College of Engineering, Peking University, Beijing 100871, People's Republic of China

Recent studies show that two low-energy Van Hove singularities (VHSs) seen as two pronounced peaks in the density of states could be induced in twisted graphene bilayer. Here, we report angle dependent VHSs of slightly twisted graphene bilayer studied by scanning tunneling microscopy and spectroscopy. We show that energy difference of the two VHSs follows $\Delta E_{vhs} \sim \hbar v_F \Delta K$ between 1.0º and 3.0º (here $v_F \sim 1.1 \times 10^6$ m/s is the Fermi velocity of monolayer graphene, $\Delta K = 2K\sin(\theta/2)$ is the shift between the corresponding Dirac points of the twisted graphene bilayer). This result indicates that the rotation angle between graphene sheets not results in significant reduction of the Fermi velocity, which quite differs from that predicted by band structure calculations. However, around a twisted angle $\theta \sim 1.3º$, the observed $\Delta E_{vhs} \sim 0.11$ eV is much smaller than the expected value $\hbar v_F \Delta K \sim 0.28$ eV at 1.3º. The origin of the reduction of $\Delta E_{vhs}$ at 1.3º is discussed.

Graphene, a two-dimensional honeycomb lattice of carbon atoms, is considered as a strong candidate for post-silicon electronic devices [1-7]. Its topological features of the electronic states can be changed by lattice deformations [8-13]. This is of fundamental importance in providing building blocks and device concepts for an all-graphene circuit in the future. Compared with single-layer graphene, bilayer graphene displays even more complex electronic band structures and intriguing properties [14-30]. Recent studies reveal that the low-energy band structure of graphene bilayer is extremely sensitive to the stacking sequence [14,18]. Two low-energy Van Hove singularities (VHSs), which originate from the two saddle points in the band structure, were observed in twisted graphene bilayer as two pronounced peaks in the density of states (DOS) [25,30]. Li et al. studied the VHSs of graphene bilayer (a chemical vapor deposition (CVD)-grown graphene monolayer deposited on a graphite surface) with three different twisted angles by scanning tunneling microscopy and spectroscopy (STM and STS). They demonstrated that the energy difference of the two VHSs show a strong angle dependence [25]. After this seminal observation, several authors addressed the physics of twisted graphene bilayer theoretically and obtained many interesting results [31-40]. The most striking results are the significant angle-dependent reduction of the Fermi velocity and the appearance of almost dispersionless bands (flat bands) around 1º ~ 1.5º [32,33,35,37-39]. It suggests that electrons in graphene bilayer can be changed from ballistic to localized by simply varying the rotation angle. However, a systematic experimental study of twisted angle dependent band structures in graphene bilayer is still scarce so far.

In this Letter, we address the twisted angle dependent VHSs in graphene bilayer (with twisted angle < 3.0º). The morphology and local DOS of the graphene bilayer with as many as eight different twisted angles were studied by STM and STS, respectively. The energy difference of the two VHSs increases linear with the sine of twisted angle, i.e., $\Delta E_{vhs} \sim \hbar v_F \Delta K$, between 1.0º and 3.0º. Here $v_F \sim 1.1 \times 10^6$ m/s, $\Delta K = 2K\sin(\theta/2)$ the shift between the corresponding Dirac points of the twisted graphene bilayer, and $K = 4\pi/3a$ (a ~ 0.246 nm is the lattice constant of the hexagonal lattice). Our result indicates that the rotation angle between graphene sheets not results in the reduction of the Fermi velocity. This differs much from that predicted by theory.

The graphene bilayer was grown on a 25 micron thin polycrystalline Rh foil via a traditional ambient pressure CVD method. The process is similar to the systhesis of graphene on Pt and Cu foils, which were reported in previous papers [41,42]. The details of synthesis of the sample is described in the supplemental material [43]. The thickness of the as-grown graphene was characterized by Raman spectra measurements (see Fig. S1 in the supplemental material [43]), as reported in our previous papers [41,42]. The STM system was an ultrahigh vacuum four-probe scanning probe microscope from UNISOKU. All STM and STS measurements were performed at liquid-nitrogen temperature and the images were taken in a constant-current scanning mode. The STM tips were obtained by chemical etching from a wire of Pt(80%) Ir(20%) alloys. Lateral dimensions observed in the STM images were calibrated using a standard graphene lattice. The STS spectrum, i.e., the dI/dV-V curve, was carried out with a standard lock-in technique using a 957 Hz alternating current modulation of the bias voltage.

Figure 1(a) shows a large-area STM image of the graphene grown on polycrystalline Rh foil taken from a flat terrace of Rh surface (see Fig. S2 in the supplemental material [43] for a larger STM image). Clear periodic protuberances with a period of 4.9 nm are observed, as shown in Fig. 1(b). The periodic protuberances are attributed to the Moiré pattern arising from a stacking

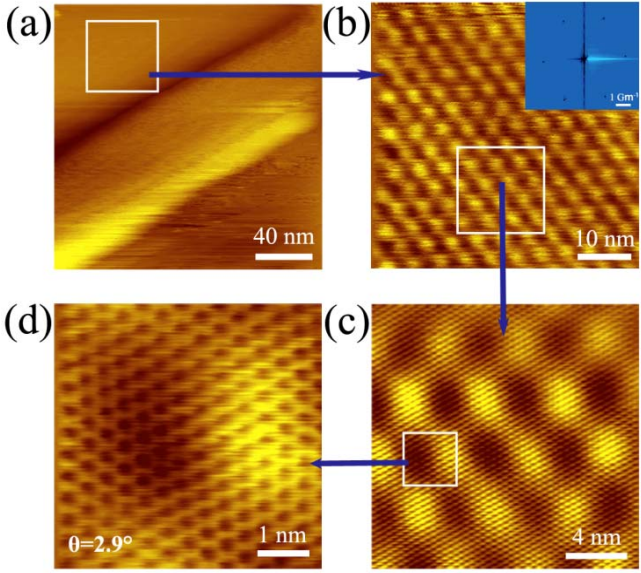

FIG. 1 (color online). (a) A 200×200 nm$^2$ STM image of graphene on Rh foil ($V_{sample}$ = -600 mV and I = 0.34 nA). (b) Zoom-in topography of the white frame in (a) shows a moiré pattern with a period of 4.9 nm ($V_{sample}$ = -280 mV and I = 0.06 nA). The inset is Fourier transforms of the superstructures. (c) Zoom-in image of white frame in panel (b) ($V_{sample}$ = -351 mV and I = 0.14 nA). (d) Atomic-resolution image of the white frame in panel (c) ($V_{sample}$ = -496 mV and I = 0.17 nA). The twisted angle of the graphene bilayer is estimated as about 2.9°.

misorientation between the top graphene layer and the underlaying layer. The twisted angle θ is related to the period of the Moiré pattern by D = a/(2sin(θ/2)) and is estimated as 2.9°. For monolayer graphene on a (111) surface of single-crystal Rh, the lattice mismatch between graphene (0.246 nm) and Rh(111) (0.269 nm) and the strong C-Rh covalent bond also lead to hexagonal Moiré superstructures. However, the expected periodicity is only about 2.9 nm resulting from a 12C/11Rh coincidence lattice [44-46] (see Fig. S3 in the supplemental material [43] for STM images of monolayer graphene on a (111) surface of single-crystal Rh.). The smallest period of the Moiré pattern studied in this Letter is about 4.9 nm (corresponding to the twisted angle 2.9°). This eliminates the lattice mismatch of monolayer graphene and Rh(111) as the origin of the observed periodic protuberances. Additionally, for monolayer graphene grown on polycrystalline Rh foil, the coupling between graphene and the substrate is much weaker than that of monolayer graphene on a single-crystal Rh and the coupling strength varies on different monolayer graphene (see Fig. S4 in the supplemental material [43] for two typical STS curves recorded on two different monolayer graphene on polycrystalline Rh foil.). Due to the weak coupling, no periodic Moiré superstructures can be seen in monolayer graphene grown on polycrystalline Rh foil.

Figure 1(c) and (d) show the atomic resolution STM images of the graphene. In twisted graphene bilayer, there is local Bernal-stacked regions, where a triangular lattice is expected to be seen [47]. However, only honeycomb lattice can be observed in atomic resolution STM image of the sample both on and between the protuberances, as shown in Fig. 1(d). In literature [47,48], both the triangular and the honeycomb lattice have been reported by many groups in graphite with Bernal-stacked lattice. The exact origin of this phenomenon is still not clear, as discussed in Ref. 47.

Graphene bilayer with different twisted angles can be observed on the graphene samples grown on Rh foils. Figure 2 shows six typical STM topographs of graphene bilayer with different stacking misorientation angles. These systems provide an ideal platform to study the twisted angle dependent VHSs. For twisted graphene bilayer, the Dirac points of the two layers no longer coincide and the zero-energy states occur at k = -$\Delta K$/2 in layer 1 and k = $\Delta K$/2 in layer 2. The displaced Dirac cones cross at energies $\pm\hbar v_F \Delta K/2$ and two saddle points are unavoidable along the intersection of the two cones when there is a finite interlayer hopping [18].

Figure 3 (a) and (b) show the STM image and STS

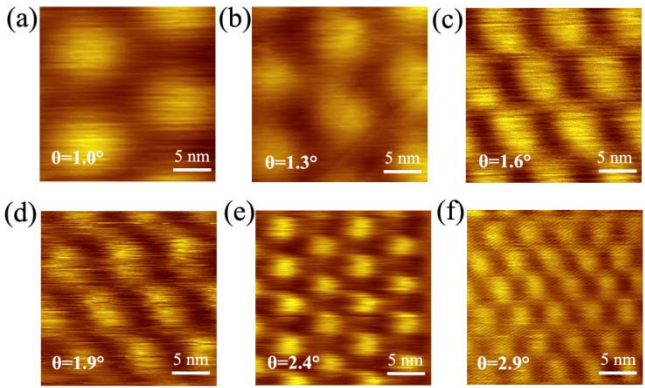

FIG. 2 (color online). (a-f) The 22 nm × 22 nm STM topographs of graphene bilayer with six different twisted angles. The period of the moiré pattern changes as a function of the twisted angle θ. (a) θ = 1.0°, D = 14.1 nm, $V_{sample}$ = 317 mV, I = 0.47 nA. (b) θ = 1.3°, D = 10.8 nm, $V_{sample}$ = 560 mV, I = 0.33 nA. (c) θ = 1.6°, D = 8.8 nm, $V_{sample}$ = 246 mV, I = 0.88 nA. (d) θ = 1.9°, D = 7.4 nm, $V_{sample}$ = -375 mV, I = 0.11 nA. (e) θ = 2.4°, D = 5.8 nm, $V_{sample}$ = -304 mV, I = 0.34 nA. (f) θ = 2.9°, D = 4.9 nm, $V_{sample}$ = -600 mV, I = 0.17 nA.

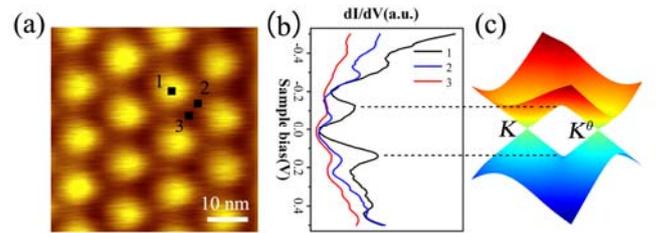

FIG. 3 (color online). (a) A typical STM iamge of graphene bilayer with the twisted angle ~1.1° ($V_{sample}$ = -317 mV and I = 0.47 nA). The period of the moiré pattern is about 12.8 nm. (b) Tunneling spectra recorded on bright and dark regions of the moiré pattern at positions indicated in panel (a). The spectra show two peaks attributing to two Van Hove singularities. (c) Electronic band structure of twisted bilayer graphene with a finite interlayer coupling. Two saddle points (VHSs) form at k = 0 between the two Dirac cones, K and $K_\theta$, with a separation of $\Delta K = 2K\sin(\theta/2)$. The low-energy VHSs contribute to two pronounced peaks flanking zero-bias in the tunneling spectra.

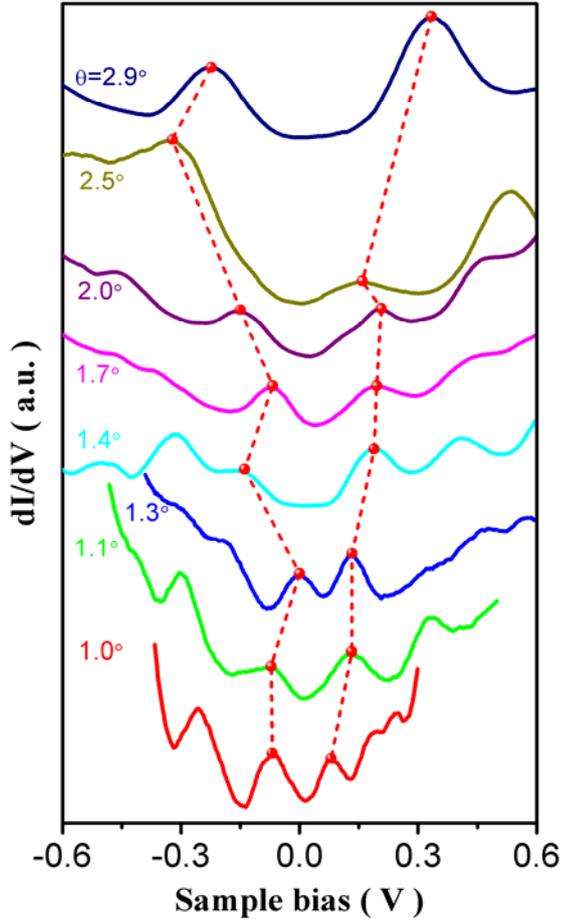

FIG. 4 (color online). Eight dI/dV-V curves taken on graphene bilayer with different twisted angles. Center of the two peaks flanking zero bias, which originate from the two Van Hove peaks of local density of states, are indicated by red solid dots.

spectra of graphene bilayer with a twisted angle ~ 1.1°, respectively. Although the peak heights and degree of asymmetry of the spectra depend on their positions in the Moiré pattern (bright or dark regions in Fig. 3(a)), all STS spectra show two peaks flanking zero-bias, as shown in Fig. 3(b). Similar position dependent spectra were also observed in the Moiré pattern of the CVD-grown graphene deposited on a graphite surface [25]. The tunnelling spectrum gives direct access to the local DOS (LDOS) of the surface at the position of the STM tip. The two peaks in the tunneling spectra are attributed to the two Van Hove peaks in DOS, which originate from the two saddle points of the band structure, as shown in Fig. 3(c) (see Fig. S5 and the supplemental material [43] for details of analysis).

Figure 4 shows eight typical tunneling curves taken on graphene bilayer with different twisted angles. In order to ascertain the reproducibility of the results, several tens of tunneling spectra on different positions of the graphene bilayer with different twisted angles are recorded. Although the peak heights and peak positions of the spectra vary slightly, the main features of these dI/dV-V curves are almost completely reproducible (for example, see Fig. S6 of the supplemental material [43] for more STS spectra). Obviously, the two VHSs of these samples show a strong angle dependent energy difference. At a twisted angle $\theta \sim 1.3°$, the two tunneling peaks show the least energy difference, which will be discussed subsequently. Additionally, the positions of the two VHSs are not always symmetric about the Fermi level, suggesting charge transfer between the graphene and the substrate. The magnitude of the charge transfer should mainly depend on the coupling strength between the sublayer graphene and the substrate, which varies in different samples (see Fig. S4 in the supplemental material [43]).

Figure 5 summarizes the energy difference of the two VHSs $\Delta E_{vhs}$ as a function of the twisted angles (see Fig. S7 of the supplemental material [43] for methods to choose the positions of VHSs.). Except at 1.3°, $\Delta E_{vhs}$ increase linear with the sine of the twisted angle (or the twisted angle for small angles). Theoretically, it is predicted that the position of the two VHSs can be simply estimated by $\Delta E_{vhs} = \hbar v_F' \Delta K - 2t_\theta$ [25,33,37]. Here, $t_\theta$ is the interlayer hopping parameter and $v_F'$ the renormalized Fermi velocity of bilayer graphene. With assuming $t_\theta$ is a constant that independent of the twisted angle, the value of $\Delta E_{vhs}$ is expected to increase linear with the twisted angle. Li et al. studied the VHSs of the CVD-grown graphene monolayer deposited on graphite with three different twisted angles and reported this linear dependence (in their experiment, $2t_\theta \sim 0.216$ eV is obtained.) [25]. Their result is also plotted along with our experimental data in Fig. 5. Obvious deviation between our experimental result and their data is observed.

Our experimental result reveals that the energy difference of the two VHSs follows $\Delta E_{vhs} \sim \hbar v_F \Delta K$

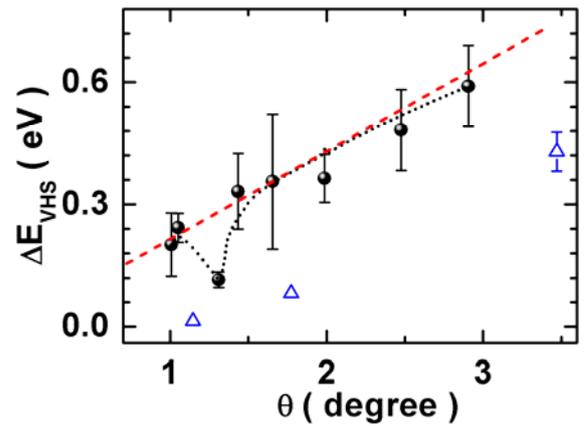

FIG. 5 (color online). The energy difference of the two VHSs $\Delta E_{vhs}$ as a function of the twisted angles. The solid black circles are the average experimentally measured values obtained from several tens of tunneling spectra for each twisted angle and the error bars in energy represent the minimum and maximum observed energies of $\Delta E_{vhs}$. The open triangles are the experimental results taken from the CVD-grown graphene sheet deposited on graphite, as reported in [25]. The red dashed line is plotted according $\Delta E_{vhs} = \hbar v_F \Delta K$, where $v_F \sim 1.1 \times 10^6$ m/s, $\Delta K = 2K\sin(\theta/2)$). The black dotted curve is the guide to eyes.

between 1.0º and 3.0º except at 1.3º, where $v_F \sim 1.1 \times 10^6$ m/s. It suggests that the displaced Dirac cones of slightly twisted graphene bilayer cross and two saddle points are formed along the intersection of the two cones at energies about $\pm \hbar v_F \Delta K/2$. This indicates that the rotation angle between graphene sheets not results in significant reduction of the Fermi velocity. In literature, Landau-level spectroscopy of twisted graphene bilayer demonstrated a negligible renormalization of the Fermi velocity [49]. Hicks et al. measured the band structure of slightly twisted graphene bilayer with θ ~ 1.7º, 2.4º, and 4.2º by using angle-resolved photoemission and also did not detect any significant change of the Fermi velocity [27]. Very recently, the problem about the Fermi velocity of slightly twisted graphene bilayer is discussed theoretically and two distinct electronic states of this coupled system are predicted to account for the two distinct results [50]. In refs. [25,26], the experiment was carried out on CVD-grown graphene monolayer deposited on graphite and the top layer of graphite was considered as an isolated monolayer graphene. In fact, there is strong coupling between the top layer and the deeper layers below the graphite surface. The electronic band structure of a graphite differs quite from that of monolayer graphene. This may be the origin of the observed reduction of Fermi velocity in [25,26].

The large deviation of $\Delta E_{vhs}$ between our experimental result and the data in Ref.[25] mainly arises from the magnitude of the interlayer hopping parameter. In their experiment, $2t_\theta \sim 0.216$ eV is obtained to account for the experimental result. In our experiment, the two saddle points appear at energies about $\pm \hbar v_F \Delta K/2$, which suggests that the magnitude of $t_\theta$ is negligible compared with $\hbar v_F \Delta K$ (it is interesting to note that although the magnitude of $t_\theta$ is negligible in our system, it still results in two saddle points that lead to two VHSs in DOS.). The effect of the substrate may be the possible origin of the different interlayer hopping. Further experiments carried out on graphene bilayer grown on different substrates are expected to uncover the exact nature of this difference.

Additionally, the abrupt reduction of $\Delta E_{vhs}$ at θ ~ 1.3º observed in our experiment is very interesting. The observed $\Delta E_{vhs} \sim 0.11$ eV is much less than the expected value $\hbar v_F \Delta K \sim 0.28$ eV at 1.3º. There are two possible origins for this observation. The first one is that some unknown effects enhance the interlayer coupling strength at θ ~ 1.3º, resulting in a small $\Delta E_{vhs}$. The second non-trivial one is that there are flat bands in twisted graphene bilayer with twisted angle around 1.3º and the reduction of Fermi velocity around 1.3º leads to the small $\Delta E_{vhs}$. If this is the case, then the abrupt reduction of $\Delta E_{vhs}$ at θ ~ 1.3º is beyond the description of any continuum model [32,33,35,37-39]. A resolution of this issue requires further theoretical analysis and experiments. Very recently, two theoretical studies suggest that superconductivity could be induced in graphene by involving repulsive electron-electron interactions [51,52]. One possible approach is to raise the Fermi level up to the vicinity of a saddle-point in graphene's electronic structure [52]. This is very difficult to achieve in monolayer graphene, but to some extent easy to realize in twisted graphene bilayer, in which the saddle-point locates not far from the Fermi level, as shown in Fig. 4. For the twisted graphene bilayer with θ ~ 1.3º, one saddle-point is natural placed at the Fermi energy, as shown in Fig. 4 and Fig. S6. Further experiments will be carried out to explore the novel superconductivity in graphene bilayer.

In summary, we address the twisted angle dependent VHSs in graphene bilayer grown on Rh foil. The energy difference of the two VHSs follows $\Delta E_{vhs} \sim \hbar v_F \Delta K$, between 1.0º and 3.0º. Our result indicates that the rotation angle between graphene sheets not results in significant reduction of the Fermi velocity. The experimental results reported here suggest that the bilayer graphene grown on Rh foil provides an ideal platform for VHSs engineering of electronic properties and exploring many attractive phases.

This work was supported by the National Natural Science Foundation of China (Grant Nos. 11004010, 10804010, 10974019, 21073003, 51172029 and 91121012), the Fundamental Research Funds for the Central Universities, and the Ministry of Science and Technology of China (Grants Nos. 2011CB921903, 2012CB921404).

§ These authors contributed equally to this paper.
*Email: yanfengzhang@pku.edu.cn
      helin@bnu.edu.cn.

[1] K. S. Novoselov, A. K. Geim, S. V. Morozov, D. Jiang, Y. Zhang, S. V. Dubonos, I. V. Grigorieva, and A. A. Firsov, Science **306**, 666 (2004).
[2] A. K. Geim and K. S. Novoselov, Nature Mater. **6**, 183 (2007).
[3] A. H. Castro Neto, F. Guinea, N. M. R. Peres, K. S. Novoselov, and A. K. Geim, Rev. Mod. Phys. **81**, 109 (2009).
[4] S. Das Sarma, S. Adam, E. H. Hwang, and E. Rossi, Rev. Mod. Phys. **83**, 407 (2011).
[5] K. S. Novoselov, A. K. Geim, S. V. Morozov, D. Jiang, M. I. Katsnelson, I. V. Grigorieva, S. V. Dubonos, and A. A. Firsov, Nature **438,** 197 (2005).
[6] Y. B. Zhang , Y. W. Tan , H. L. Stormer, and P. Kim, Nature **438** , 201 (2005).
[7] M. O. Goerbig, Rev. Mod. Phys. **83**, 1193 (2011).
[8] N. Levy, S. A. Burke, K. L. Meaker, M. Panlasigui, A. Zettl, F. Guinea, A. H. Castro Neto, M. F. Crommie, Science **329**, 544 (2010).
[9] H. Yan, Y. Sun, L. He, J. C. Nie, and M. H. W. Chan, Phys. Rev. B **85**, 035422 (2012).
[10] M. M. Ugeda, I. Brihuega, F. Guinea, and J. M. Gomez-Rodriguez, Phys. Rev. Lett. **104**, 096804 (2010).
[11] M. M. Ugeda, D. Fernandez-Torre, I. Brihuega, P. Pou, A. J. Martinez-Galera, R. Perez, and J. M. Gomez-Rodriguez, Phys. Rev. Lett. **107**, 116803 (2011).
[12] L. Tarruell, D. Greif, T. Uehlinger, G. Jotzu, and T. Esslinger, Nature **483**, 302 (2012).
[13] K. K. Gomes, W. Mar, W. Ko, F. Guinea, and H. C. Manaharan, Nature **483**, 306 (2012).
[14] E. McCann and V. I. Falko, Phys. Rev. Lett. **96**, 086805 (2006).
[15] K. S. Novoselov, E. McCann, S. V. Morozov, V. I. Falko, M. I. Katsnelson, U. Zeitler, D. Jiang, F. schedin, and A. K. Geim, Nature Phys. **2**, 177 (2006).
[16] G. M. Rutter, S. Jung, N. N. Klimov, D. B. Newell, N. B. Zhitenev, and J. A. Stroscio, Nature Phys. **7** , 649 (2011).

[17] Z. Q. Li, E. A. Henriksen, Z. Jiang, Z. Hao, M. C. Martin, P. Kim, H. L. Stormer, and D. N. Basov, Phys. Rev. Lett. **102**, 037403 (2009).
[18] J. M. B. Lopes dos Santos, N. M. R. Peres, and A. H. Castro Neto, Phys. Rev. Lett. **99**, 256802 (2007).
[19] S. Kim, K. Lee, and E. Tutuc, Phys. Rev. Lett. **107**, 016803 (2011).
[20] M. Killi, S. Wu, and A. Paramekanti, Phys. Rev. Lett. **107**, 086801 (2011).
[21] R. Nandkishore and L. Levitov, Phys. Rev. Lett. **107**, 097402 (2011).
[22] N. Gu, M. Rudner, and L. Levitov, Phys. Rev. Lett. **107**, 156603 (2011).
[23] D. S. Lee, C. Riedl, T. Beringer, A. H. Castro Neto, K. von Klitzing, U. Starke, and Jurgen H. Smet, Phys. Rev. Lett. **107**, 216602 (2011).
[24] A. S. Mayorov, D. C. Elias, M. Mucha-Kruczynski, R. V. Gorbachev, T. Tudorovskiy, A. Zhukov, S. V. Morozov, M. I. Katsnelson, V. I. Falko, A. K. Geim, K. S. Novoselov, Science **333**, 860 (2011).
[25] G. H. Li, A. Luican, J. M. B. Lopes dos Santos, A. H. Castro Neto, A. Reina, J. Kong and E. Y. Andrei, Nature Phys. **6**, 109 (2010).
[26] A. Luican, G. H. Li, A. Reina, J. Kong, R. R. Nair, K. S. Novoselov, A. K. Geim, and E.Y. Andrei, Phys. Rev. Lett. **106**, 126802 (2011).
[27] J. Hicks, M. Sprinkle, K. Shepperd, F. Wang, A. Tejeda, A. Taleb-Ibrahimi, F. Bertran, P. Le Fevre, W. A. de Heer, C. Berger, and E. H. Conrad, Phys. Rev. B **83**,205403 (2011).
[28] D. L. Miller, K. D. Kubista, G. M. Rutter, M. Ruan, W. A. de Heer, M. Kindermann, P. N. First, and J. A. Stroscio, Nature Phys. **6**, 811 (2010).
[29] G. M. Rutter, S. Jung, N. N. Klimov, D. B. Newell, N. B. Zhitenev, and J. A. Stroscio, Nature Phys. **7**, 649 (2011).
[30] L. Meng, Z.-D. Chu, Y. Zhang, J.-Y. Yang, R.-F. Dou, J.-C. Nie, and L. He, Phys. Rev. B **85**, 235453 (2012).
[31] R. de Gail, M. O. Goerbig, F. Guinea, G. Montambaux, and A. H. Castro Neto, Phys. Rev. B **84**, 045436 (2011).
[32] E. Suarez Morell, J. D. Correa, P. Vargas, M. Pacheco, and Z. Barticevic, Phys. Rev. B **82**, 121407(R) (2010).
[33] R. Bistritzer and Allan H. MacDonald, Proc. Natl. Acad. Sci. U.S.A. **108**, 12233 (2011).
[34] S. Shallcross, S. Sharma, E. Kandelaki, and O. A. Pankratov, Phys. Rev. B **81**, 165105 (2010).
[35] G. T. de Laissardiere, D. Mayou, and L. Magaud, Nano Lett. **10**, 804 (2010).
[36] R. Bistritzer and A. H. MacDonald, Phys. Rev. B **84**, 035440 (2011).
[37] J. M. B. Lopes dos Santos, N. M. R. Peres, and A. H. Castro Neto, arXiv: 1202.1088.
[38] G. T. de Laissardiere, D. Mayou, and L. Magaud, arXiv: 1203.3144.
[39] P. Moon and M. Koshino, Phys. Rev. B **85**, 195458 (2012).
[40] M. Kindermann and P. N. First, arXiv: 1204.0814.
[41] T. Gao, S. Xie, Y. Gao, M. Liu, Y. Chen, Y. Zhang, and Z. Liu, ACS Nano. **5**, 9194 (2011).
[42] Y. Zhang, T. Gao, Y. Gao, S. Xie, Q. Ji, K. Yan, H. Peng, and Z. Liu, ACS Nano. **5**, 4014 (2012).
[43] See supplementary material for Raman spectra, more STM images, STS spectra, and the detail of analysis.
[44] B. Wang, M. Caffio, C. Bromley, H. Fruchtl, and R. Schaub, ACS Nano. **4**, 5773 (2010).
[45] M. Sicot, S. Bouvron, O. Zander, U. Rudiger, Yu. S. Dedkov, and M. Fonin, Appl. Phys. Lett. **96**, 093115 (2010).
[46] M. Sicot, P. Leicht, A. Zusan, S. Bouvron, O. Zander, M. Weser, Y. S. Dedkov, K. Horn, and M. Fonin, ACS Nano. **6**, 151 (2012).
[47] E. Y. Andrei, G. Li, and X. Du, Rep. Prog. Phys. **75**, 056501 (2012).
[48] P. Moriarty, and G. Hughes, Appl. Phys. Lett. **60**, 2338 (1992); P. J. Ouseph, T. Poothackanal, G. Mathew, Phys. Lett. A **205**, 65 (1995); J. .I. Paredes, A.M. Alonso, and J. M. D. Tascon, Carbon **39**, 476 (2001); S. Hembacher, F. Giessibl, J. Mannhart, and C. F. Quate., Proc. Natl. Acad. Sci. **100**, 12539 (2003); H.A. Mizes, S.-i. Park, and W.A. Harrison, Phys. Rev. B **36**, 4491 (1987); F. Atamny, O. Spillecke, and R. Schlogl, Phys. Chem. Chem. Phys. **1**, 4113 (1999); Y. Wang, Y. Ye, and K. Wu, Surface Science **600**, 729 (2006); P. Xu, Y. Yang, S.D. Barber, M.L. Ackerman, J.K. Schoelz, I.A. Kornev, S. Barraza-Lopez, L. Bellaiche, and P. M. Thibado, Phys. Rev. B **84**, 161409 (2011).
[49] D. M. Miller, K. D. Kubista, G. M. Rutter, W. A. de Heer, P. N. First, and J. A. Stroscio, Science, **324**, 9242 (2009).
[50] E. J. Mele, Phys. Rev. B **84**, 235439 (2011).
[51] M. V. Hosseini and M. Zareyan, Phys. Rev. Lett. **108**, 147001 (2012).
[52] R. Nandkishore, L. S. Levitov, and A. V. Chubukov, Nature Phys. **8**, 158 (2012).

# Angle Dependent Van Hove Singularities in Slightly Twisted Graphene Bilayer

Wei Yan[1,§], Mengxi Liu[2,§], Rui-Fen Dou[1], Lan Meng[1], Lei Feng[1], Zhao-Dong Chu[1], Yanfeng Zhang[2,3,*], Zhongfan Liu[2], Jia-Cai Nie[1], and Lin He[1,*]

[1] Department of Physics, Beijing Normal University, Beijing, 100875, People's Republic of China
[2] Center for Nanochemistry (CNC), College of Chemistry and Molecular Engineering, Peking University, Beijing 100871, People's Republic of China
[3] Department of Materials Science and Engineering, College of Engineering, Peking University, Beijing 100871, People's Republic of China

Method: The process is similar to the systhesis of graphene on Pt and Cu foils. The polycrystalline Rh foil, which is mainly (111) oriented according to our X-ray diffraction measurements (not shown), was firstly heated from room temperature to 1000 ºC in 45 min under an Ar flow of 850 sccm. Then the furnace was experienced a hydrogen gas flow of 50 sccm for 40 min at 1000 ºC. Finally, $CH_4$ gas was introduced with a flow ratio of 5-10 sccm, and the growth time is varied from 3 to 15 min for controlling the thickness of graphene. The as-grown sample is cooled down to room temperature and can be transferred into the ultrahigh vacuum condition for further characterizations.

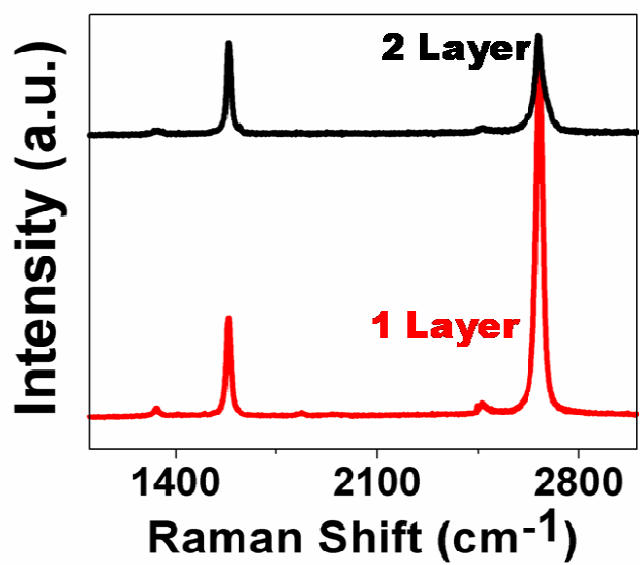

Figure. S1. Raman spectra of graphene bilayer and graphene monolayer grown on Rh foils. The synthesized graphene is transferred onto a $SiO_2$/Si substrate for Raman measurements.

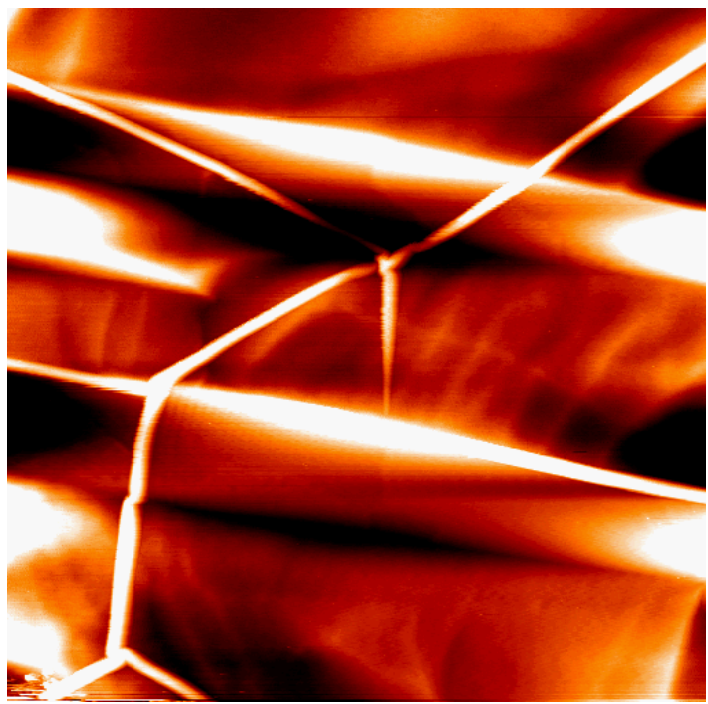

Figure S2. (1000 nm × 1000 nm; V=-0.66 V, I=3.90 nA ) Large-scale STM image showing the typical surface morphology of graphene growth on Rh foils under atmospheric growth conditions. The stripped protrusions along the terraced steps or along the Rh grain boundaries are the graphene wrinkles, usually presenting a height of several nanometers and a lateral width of several tenths of nanometers.

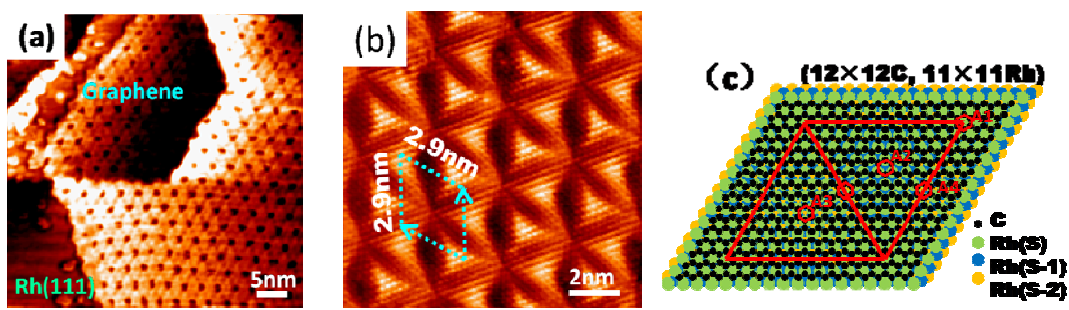

Figure S3. (a) A typical STM image of monolayer graphene grown on Rh (111) surface. It shows moiré pattern with a period of 2.9 nm. (b) Atomic-resolution image of the moiré pattern between monolayer graphene and Rh (111) surface. (c) Model of a 12C/11Rh superstructure following Ref. [S1]. A1, A2, A3, and A4 are atop, fcc, hcp, and bridge sites, respectively [S1].

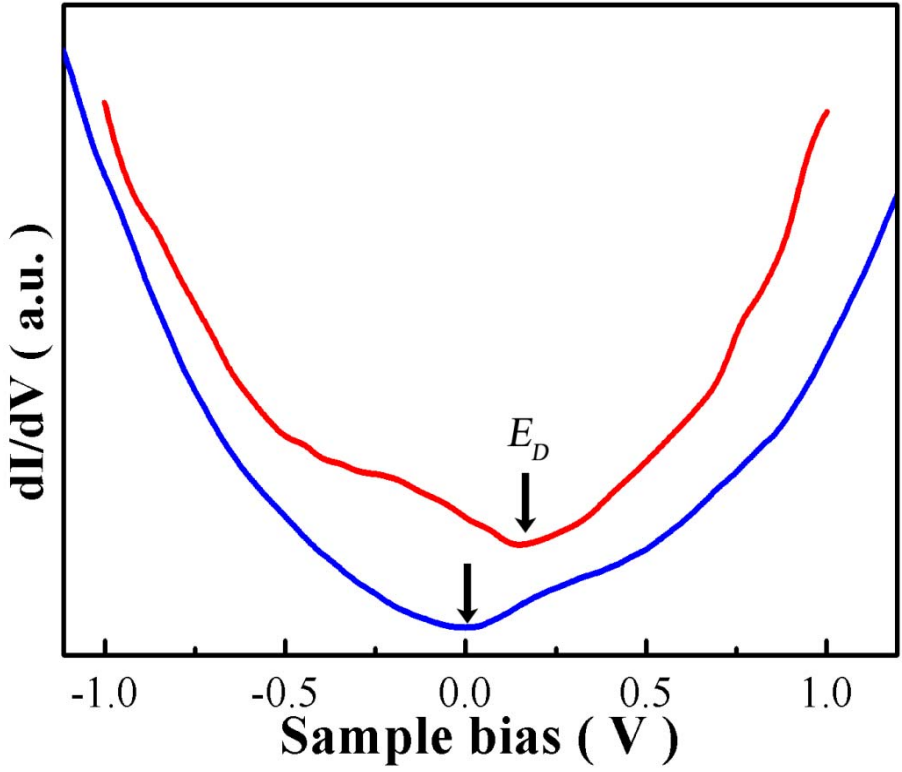

Figure S4. Two typical STS curves recorded on two different monolayer graphene on polycrystalline Rh foil. The arrows point to the positions of the Dirac points, 0.0 eV and 0.2 eV of the two different samples respectively. The magnitude of charge transfer between the graphene and the substrate, which reflects the coupling between the graphene and the substrate, varies on different monolayer graphene samples. We believe that this is the main origin of the variation of the magnitude of the charge transfer observed in twisted graphene bilayer, as shown in Fig. 4 in the main text.

**Tight binding model for twisted graphene bilayer**

The Hamiltonian for the bilayer with a twist has the form $H = H_1 + H_2 + H_\perp$, where $H_1$ and $H_2$, are the Hamiltonians for each layer, $H_\perp$ is the interaction Hamiltonian between the two layers following the model in reference S2. The Hamiltonians of $H_1$, $H_2$, and $H_\perp$ can be expressed as

$$H_1 = -t \sum_{<i,j>} (c_{A_i}^\dagger c_{B_j} + H.c.), \qquad (1)$$

$$H_2 = -t \sum_{<i,j>} (c_{A_i'}^\dagger c_{B_j'} + H.c.), \qquad (2)$$

$$H_\perp = \sum_{\alpha,\beta}^{i} t_\perp^{\alpha\beta}(r_i) c_\alpha^\dagger(r_i) c_\beta(r_i + \Delta r_i) + H.c. \quad . \tag{3}$$

Here $\Delta r_i$ is the horizontal (in-plane) displacement from an atom of layer 1 to the closest atom in layer 2, the sublattice $\alpha$ = A, B, and $\beta$ = A′, B′, $t_\perp^{\alpha\beta}(r)$ is the hopping amplitude between nearest neighbor atoms from different layers. For twisted graphene bilayer, the Dirac points of the two layers no longer coincide and the zero energy states occur at k=−ΔK/2 in layer 1 and k=ΔK/2 in layer 2, as shown in Fig. S5. For small angles of rotation, the values of $t_\perp^{\alpha\beta}(r)$ are equal for **G = 0**, **G = -G₁**, and **G = -G₁-G₂** and much smaller for all other **G** vectors. It means that the states of momentum k in layer 1 (2) are coupled directly only to states of layer 2 (1) of momentum **k** (**k**), **k+G₁**(**k-G₁**) and **k+G₁+G₂**(**k-G₁-G₂**) (here **G₁** and **G₂** are the basis vectors of the reciprocal lattice of the Moiré superlattice). As a consequence, the low energy spectrum near the Dirac point of the twisted graphene bilayer can be well described by the Hamiltonian including only the six momentum values, **k, k+G₁** and **k+G₁+G₂** in layer 2 and **k, k-G₁** and **k-G₁-G₂** in layer 1 [S2,S3].

One of the most striking results predicted by this model is the significant angle-dependent reduction of the Fermi velocity in slightly twisted graphene bilayer. However, our result indicates that the rotation angle between graphene sheets not results in significant reduction of the Fermi velocity. The experimental result reported in this Letter reveals that the energy difference of the two VHSs follows $\Delta E_{vhs} \sim \hbar v_F \Delta K$ between 1.0º and 3.0º except at 1.3º, where $v_F \sim 1.1\times10^6$ m/s is the Fermi velocity of monolayer graphene. It suggests that the displaced Dirac cones of slightly twisted graphene bilayer cross and two saddle points are formed along the intersection of the two cones at energies about $\pm\hbar v_F \Delta K/2$. It is interesting to note that although the magnitude of the interlayer hopping parameter $t_\theta$ is negligible in our system, it still results in two saddle points that lead to two VHSs in DOS.

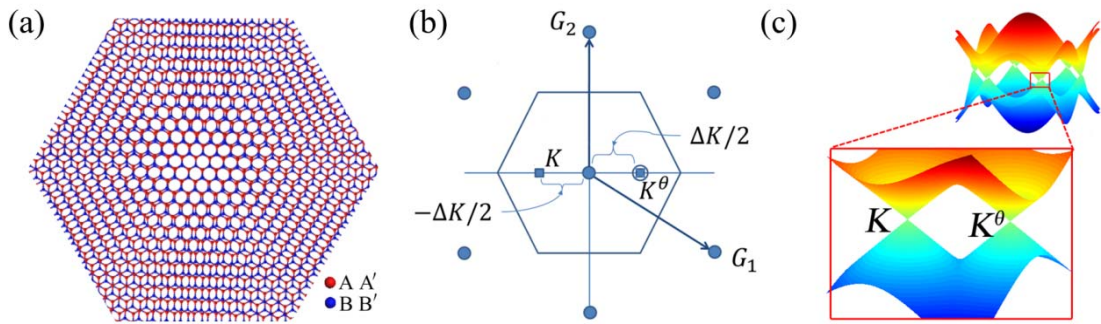

Figure S5. (a) Structural model of two misoriented honeycomb lattices with a twist angle that satisfies a condition for commensurate periodic structure leading to Moiré patterns. The two sublattices in layer 1 and 2 are denoted by A, B and A′, B′,

respectively. (b) Geometry of the Brillouin zone of the Moiré superlattice. $G_1$ and $G_2$ are the basis vectors of the reciprocal lattice of the Moiré superlattice. The zero energy states occur $-\Delta K/2$ of layer 1 and $\Delta K/2$ of layer 2. The low energy states near $\Delta K/2$ of layer 2 are coupled to states of layer 1 at $\Delta K/2$, $\Delta K/2-G_1$, and $\Delta K/2 -G_1-G_2$. Two saddle points (VHSs) form at k = 0 between the two Dirac cones, K and $K_\theta$. (c) Electronic band structure of twisted bilayer graphene with a finite interlayer coupling calculated with the four-band model. Two saddle points (VHSs) form at $k = 0$ between the two Dirac cones, $K$ and $K_\theta$, with a separation of $\Delta K = 2K\sin(\theta/2)$. The low-energy VHSs contributes to two pronounced peaks flanking zero-bias in a typical tunneling spectrum obtained in our experiment.

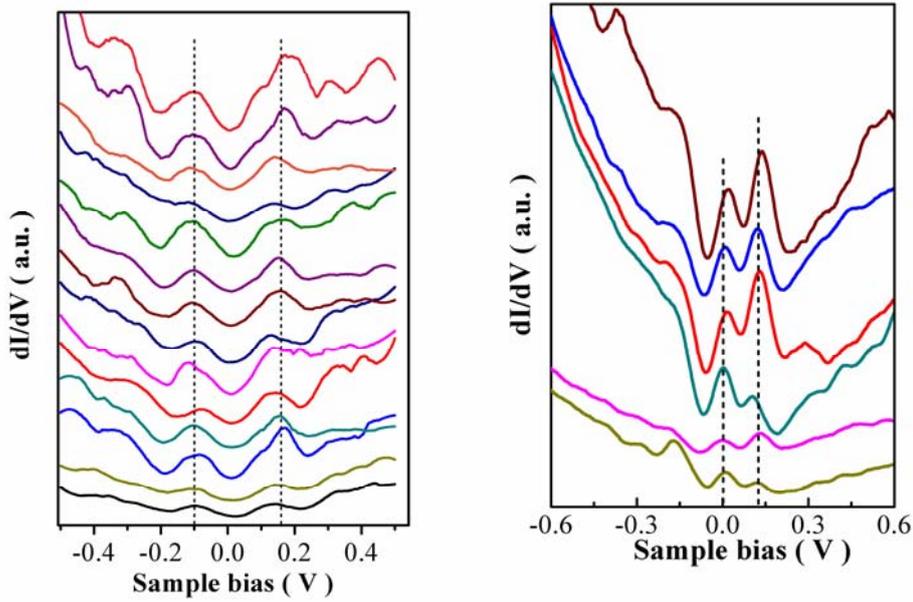

Figure S6. Left panel: fourteen dI/dV-V curves taken on graphene bilayer with the twisted angle 1.1°. Right panel: six dI/dV-V curves taken on graphene bilayer with the twisted angle 1.3°. Although the peak heights and peak positions of the spectra vary slightly, the main features of these dI/dV-V curves are almost completely reproducible. The dashed lines are guide to the eyes.

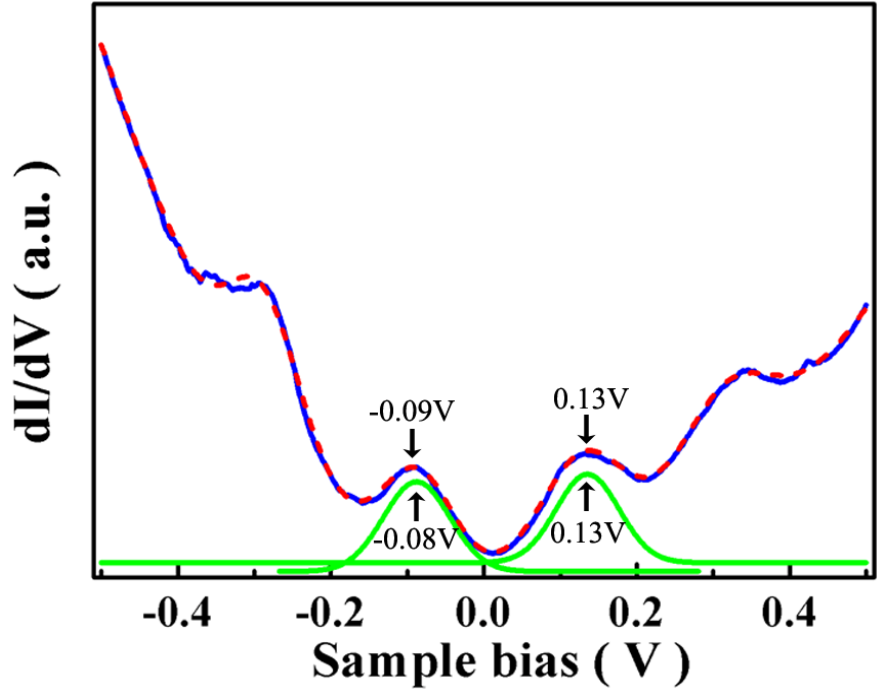

Figure S7. A typical *dI/dV-V* curve (blue solid curve) taken on graphene bilayer with the twisted angle 1.1°. The red dashed curve is the sum of two Guassian peaks (shown in green) and a simple polynomial background. The two methods result in almost identical positions of the VHSs.

[S1] B. Wang, M. Caffio, C. Bromley, H. Fruchtl, and R. Schaub, ACS Nano. **4**, 5773 (2010).
[S2] J. M. B. Lopes dos Santos, N. M. R. Peres, and A. H. Castro Neto, Phys. Rev. Lett. **99**, 256802 (2007).
[S3] G. H. Li, A. Luican, J. M. B. Lopes dos Santos, A. H. Castro Neto, A. Reina, J. Kong and E. Y. Andrei, Nature Phys. **6**, 109 (2010).